\documentclass[10pt]{article}

\usepackage{amsxtra,amssymb,amsthm,amsmath,latexsym}

\usepackage{graphicx}

\newcommand{\R}{{\mathbb R}}

\newcommand{\bee}{\begin{equation*}}
\newcommand{\eee}{\end{equation*}}
\newcommand{\be}{\begin{equation}}
\newcommand{\ee}{\end{equation}}
\newcommand{\pn}{\par\noindent}
\title{Scattering of electromagnetic waves by many thin cylinders }
\author{A G Ramm\\
\small Department of Mathematics\\[-0.8ex]
\small Kansas State University, Manhattan, KS 66506-2602, USA\\[-0.8ex]
\small \texttt{ramm@math.ksu.edu}\\
}
\date{}
\begin{document}
Results in Physis, {\bf 1}, (2011), 13-16.

\maketitle
\begin{abstract}
Electromagnetic wave scattering by many parallel to $z-$axis, thin,
perfectly conducting, circular infinite cylinders
is studied asymptotically as $a\to 0$.
Let  $D_m$ be the crossection
of the $m-$th cylinder, $a$ be its radius, and
$\hat{x}_m=(x_{m1},x_{m2})$ be its
center, $1\le m \le M$, $M=M(a)$.
It is assumed that the points $\hat{x}_m$ are distributed
so that
$$\mathcal{N}(\Delta)=\ln\frac{1}{a}\int_{\Delta}N(x)dx[1+o(1)],
$$
where $\mathcal{N}(\Delta)$ is the number of points
$\hat{x}_m$ in an arbitrary open subset $\Delta$ of the plane
$xoy$. The function $N(x)\geq 0$ is
a given continuous function. An equation for the self-consistent
(efficient) field is derived as $a\to 0$. The cylinders are assumed
perfectly conducting. A formula is derived for the effective refraction
coefficient in
the medium in which many cylinders are distributed. These cylinders may
model nanowires embedded in the medium. Our result shows how these
cylinders influence the refraction coefficient of the medium.
\end{abstract}
\pn{
 }

\section{Introduction}
There is a large literature on electromagnetic (EM) wave scattering
by an array of parallel cylinders (see, e.g., \cite{M}, where there
are many references given, and \cite{MV}).
Electromagnetic wave scattering by many parallel to $z-$axis, thin,
perfectly conducting, circular, of radius $a$, infinite cylinders
is studied in this paper asymptotically as $a\to 0$.
 The cylinders are thin in the sense $ka\ll 1$,
where $k$ is the wave number in the exterior of the cylinders,

The novel points in this paper include:

1) The solution to the wave scattering problem is considered  in
the limit $a\to 0$ when the number $M=M(a)$ of the cylinders tends
to infinity at a suitable rate. The equation for the limiting (as
$a\to 0$) effective (self-consistent) field in the medium is
derived,

2) This theory is a basis for a method for changing
refraction coefficient in a medium. The thin cylinders model nanowires
embedded in the medium. The basic physical resul of this paper is
formula  \eqref{e47}, which shows how the embedded thin cylinders change
the refraction coefficient $n^2(x)$.

Some extension of the author's
results (\cite{R515}-\cite{R595}) is obtained for EM wave scattering
by many thin perfectly conducting cylinders.

Let $D_m$, $1\leq m\leq M$, be a set of non-intersecting domains on
a plane $P$, which is $xoy$ plane. Let $\hat{x}_m\in D_m$,
$\hat{x}_m=(x_{m1},x_{m2})$, be a point inside $D_m$ and $C_m$ be
the cylinder with the cross-section $D_m$ and the axis, parallel to
$z$-axis, passing through $\hat{x}_m$. We assume that $\hat{x}_m$
is the center of the disc $D_m$ if $D_m$ is a disc of radius $a$.

Let us assume  that the cylinders are perfect conductors. Let
$a=0.5$diam$D_m$. Our basic assumptions are
\be\label{e1} ka\ll 1,
\ee where $k$ is the wave number in the region exterior to the union
of the cylinders, and \be\label{e2} \mathcal{N}(\Delta)=\ln
\frac{1}{a}\int_{\Delta}N(\hat{x})d\hat{x}[1+o(1)],\quad a\to 0, \ee
where $\mathcal{N}(\Delta)=\sum_{\hat{x}_m\in \Delta}1$ is the
number of the cylinders in an arbitrary open subset of the plane
$P$, $N(\hat{x})\geq 0$ is a continuous function, which can be
chosen as we wish. The points $\hat{x}_m$ are distributed in an arbitrary
large but fixed bounded domain on the plane $P$. We denote by
$\Omega$ the union of domains $D_m$, by $\Omega'$ its complement in
$P$, and by $D'$ the complement of $D$ in $P$. The complement in
$\R^3$ of the union $C$ of the cylinders $C_m$ we denote by $C'$.

The EM wave scattering problem consists of finding the solution to
Maxwell's equations

\be\label{e3} \nabla \times E=i\omega \mu H, \ee \be\label{e4}
\nabla \times H=-i\omega \epsilon E, \ee in $C'$, such that
\be\label{e5} E_t=0 \text{ on } \partial C, \ee where $\partial C$
is the union of the surfaces of the cylinders $C_m$, $E_t$ is the
tangential component of $E$, $\mu$ and $\epsilon$ are constants in
$C'$, $\omega$ is the frequency, $k^2=\omega^2\epsilon \mu$, $k$ is
the wave number. Denote by $n_0^2=\epsilon \mu$, so $k^2=\omega^2
n_0^2$. The solution to \eqref{e3}-\eqref{e5} must have the
following form \be\label{e6} E(x)=E_0(x)+v(x),\quad
x=(x_1,x_2,x_3)=(x,y,z)=(\hat{x},z), \ee where $E_0(x)$ is the
incident field, and $v$ is the scattered field satisfying the
radiation condition \be\label{e7} \sqrt{r}\left(\frac{\partial
v}{\partial r}-ikv\right)=o(1),\quad r=(x_1^2+x_2^2)^{1/2}, \ee and
we assume that 
\be\label{e8} E_0(x)=k^{-1}e^{i\kappa y+i k_3z}(-k_3e_2+\kappa e_3),\quad
\kappa^2+k_3^2=k^2, \ee $\{e_j\}$, $j=1,2,3$, are the unit vectors
along the Cartesian coordinate axes $x,y,z$. 
We consider EM waves
with $H_3:=H_z=0$, i.e., E-waves, or TH waves, \be\label{e9}
E=\sum_{j=1}^3E_je_j,\quad H=H_1e_1+H_2e_2=\frac{\nabla\times
E}{i\omega \mu}. \ee One can prove (see Appendix)  that the
components of $E$ can be expressed by the formulas: \be\label{e10}
E_j=\frac{ik_3}{\kappa^2}U_{x_j}e^{ik_3z},\quad j=1,2,\quad
E_3=Ue^{ik_3z},\quad U=\frac {\kappa}{k}u,  \ee where  
$u_{x_j}:=\frac{\partial u}{\partial
x_j}$, $u=u(x,y)$ solves the problem \be\label{e11}
(\Delta_2+\kappa^2)u=0\text{ in }\Omega' \ee \be\label{e12}
u|_{\partial \Omega}=0, \ee \be\label{e13} u=e^{i\kappa y}+w, \ee
and $w$ satisfies the radiation condition \eqref{e7}. Similar
calculations are done with Borgnis potentials (see, e.g.,
\cite{LL}). The unique solution to \eqref{e3}-\eqref{e8} is given by
the formulas: \be\label{e14} E_1=\frac{ik_3}{\kappa^2}U_xe^{ik_3z}
,\quad E_2=\frac{ik_3}{\kappa^2}U_ye^{ik_3z},\quad
E_3=Ue^{ik_3z},\ee \be\label{e15} H_1=\frac{k^2}{i\omega \mu
\kappa^2}U_ye^{ik_3z},\quad H_2=-\frac{k^2}{i\omega \mu
\kappa^2}U_x e^{ik_3 z},\quad H_3=0, \ee where $U_x:=\frac{\partial
U}{\partial x}$, $U_y$ is defined similarly, and
$u=u(\hat{x})=u(x,y)$ solves scalar two-dimensional problem
\eqref{e11}-\eqref{e13}.  These formulas are derived in the Appendix
for convenience of the reader.

Problem \eqref{e11}-\eqref{e13} has
a unique solution (see, e.g., \cite{R190}).

Our goal is to derive an asymptotic
formula for this solution as $a\to 0$. Our results
include formulas for the solution to the scattering problem,
derivation of the equation for the effective field in the medium
obtained by embedding many thin perfectly conducting cylinders,
and a formula for the refraction coefficient in this limiting medium.
This formula shows that by choosing suitable distribution of the
cylinders, one can change the refraction coefficient, one can make
it smaller than the original one.

The paper is organized as follows.

In Section 2 we derive an
asymptotic formula for the solution to \eqref{e11}-\eqref{e13} when
$M=1$, i.e., for scattering by one cylinder.

In Section 3 we derive
a linear algebraic system for finding some numbers that define the
solution to problem \eqref{e11}-\eqref{e13} with $M>1$. Also in
Section 3 we derive an integral equation for the effective
(self-consistent) field in the medium with $M(a)\to \infty$
cylinders as $a\to 0$. At the end of Section 3 these results are
applied to the problem of changing the
refraction coefficient of a given material by embedding many
thin perfectly conducting cylinders into it.

In Section 4 conclusions are formulated.

In Appendix formulas \eqref{e14}-\eqref{e15} are derived.

\section{EM wave scattering by one thin perfectly conducting cylinder}
Consider problem \eqref{e11}-\eqref{e13} with $\Omega=D_1$,
$\Omega'$ being the complement to $D_1$ in $\R^2$. Assume for simplicity
that
$D_1$ is a circle $x_1^2+x_2^2\leq a^2$.

Let us look for a solution
of the form \be\label{e16} u=e^{i\kappa
y}+\int_{S_1}g(\hat{x},t)\sigma(t)dt,\qquad
g(\hat{x},t):=\frac{i}{4}H_0^{(1)}(\kappa|\hat{x}-t|), \ee where
$S_1$ is the boundary of $D_1$, $H_0^{(1)}$ is the Hankel function
of order $1$, with index $0$,  and $\sigma$ is to be found from the
boundary condition \eqref{e12}. It is known (see, e.g., \cite{R476}) that
\be\label{e17} g(\kappa r)=\alpha(\kappa)+\frac{1}{2\pi}\ln
\frac{1}{r}+o(1),\,\,\,\text{ as }\,\,\, r\to 0, \ee where 
\be\label{e18}
\alpha(\kappa):=\frac{i}{4}+\frac{1}{2\pi}\ln
\frac{2}{\kappa}. \ee 
The function \eqref{e16} satisfies equations
\eqref{e11} and \eqref{e13} for any $\sigma$, and if $\sigma$ is
such that function \eqref{e16} satisfies boundary condition
\eqref{e12}, then $u$ solves problem \eqref{e11}-\eqref{e13}.
We assume $\sigma$ sufficiently smooth (H\"older-continuous is
sufficient).

The
solution to problem \eqref{e11}-\eqref{e13} is known to be unique
(see, e.g., \cite{R190}). Boundary condition \eqref{e12} yields
\be\label{e19} -u_0(s)=\alpha(\kappa)Q+\int_{S_1}g_0(s,t)
\sigma(t)dt,\qquad Q:=\int_{S_1}\sigma(t)dt, \ee \be\label{e20}
u_0(s):=e^{i\kappa s_2}, \ s\in S_1;\,\,\,
g_0(s,t)=\frac{1}{2\pi}\ln\frac{1}{r_{st}},\,\, r_{st}:=|s-t|.\ee If
$ka\ll 1$, $k^2=\kappa^2+k_3^2$, then
$$u_0(s)=1+O(\kappa a).$$

Equation \eqref{e19} is uniquely solvable for $\sigma$ if $a$ is
sufficiently small \cite{R476}.

We are interested in finding
asymptotics of $Q$ as $a\to 0$, because $u(\hat{x})$ in \eqref{e16}
can be well approximated in the region $|\hat{x}|\gg a$ by the
formula \be\label{e21}\begin{split}
u(\hat{x})&=u_0(\hat{x})+g(\hat{x},0)Q+o(1),\quad a\to 0,\\
u_0(\hat{x})&=e^{i\kappa x_2}, \quad x_2=y.
\end{split}\ee
To find asymptotics of $Q$ as $a\to 0$, let us integrate
equation \eqref{e19}
over $S_1$ and obtain \be\label{e22}
-u_0(0)|S_1|=\alpha(\kappa)Q|S_1|-\int_{S_1}dt
\sigma(t)\frac{1}{2\pi}\int_{S_1}\ln r_{st}ds, \ee where $|S_1|$ is
the length of $S_1$, $|S_1|=2\pi a$ if $S_1$ is the circle
$|\hat{x}|=a$, and $ r_{st}=|s-t|$. Denote \be\label{e23}
I:=\frac{1}{2\pi}\int_{S_1}\ln r_{st}dt=O(a\ln a),\quad a\to 0. \ee
If $S_1$ is the circle $|\hat{x}|=a$, integral \eqref{e23} can be
calculated analytically:
\be\label{e24}\begin{split}
I&=\frac{a}{2\pi}\int_0^{2\pi}\ln
\sqrt{2a^2-2a^2\cos(\psi-\varphi)}d\varphi\\
&=\frac{a}{2\pi}\int_0^{2\pi}\ln\sqrt{2a^2}d\varphi+\frac{a}{2\pi}\int_0^{2\pi}\ln\sqrt{2\sin^2\frac{\psi-\varphi}{2}}d\varphi.
\end{split}\ee
Thus, \be\label{e24a} I=a\ln(\sqrt{2}a)+\frac{a\ln
2}{2}+\frac{a}{2\pi}\int_0^{2\pi}\ln
|\sin\frac{\psi-\varphi}{2}|d\varphi. \ee One can derive that
\be\label{e25} \int_0^{2\pi}\ln
|\sin\frac{\psi-\varphi}{2}|d\varphi=2\int_0^{\pi}\ln|\sin
\theta|d\theta=-2\pi \ln 2. \ee
Indeed, if
$J:=\int_0^\pi\ln|\sin\theta|d\theta$, then \bee
J=\int_0^\pi\ln2|\sin\frac{\theta}{2} \cos\frac{\theta}{2}|d\theta=\pi
\ln 2+\int_0^\pi \ln \sin\frac{\theta}{2}d\theta+\int_0^\pi \ln
|\cos\frac{\theta}{2}| d\theta=\pi\ln 2+2J. \eee
Thus, $J=-\pi \ln 2$.
From \eqref{e24} and
\eqref{e25} one gets \be\label{e26} I=a\ln
a(1+O\left(\frac{1}{|\ln a|}\right),\quad a\to 0. \ee From \eqref{e22}
and
\eqref{e26} it follows that
\be\label{e27} Q=-\frac{2\pi u_0(0)}{\ln
\frac{1}{a}}[1+O\left(\frac{1}{|\ln a|}\right)],\quad a\to 0. \ee
Therefore, the asymptotic
solution to the scattering problem \eqref{e11}-\eqref{e13} in the
case of one circular cylinder of radius $a$, as $a\to 0$, is
\be\label{e28} u(\hat{x})\sim u_0(\hat{x})-\frac{2\pi}{\ln
\frac{1}{a}}g(\hat{x},0)u_0(0),\,\,\, a\to 0,\,\,\, |\hat{x}|>a. \ee
Electromagnetic wave, scattered by the single cylinder, is
calculated by formulas \eqref{e14}-\eqref{e15} in which
$u=u(\hat{x}):=u(x_1,x_2)$ is given by formula \eqref{e28}.

\section{Wave scattering by many thin cylinders}
Problem \eqref{e11}-\eqref{e13} should be solved  when
$\Omega$ is a union of many small domains $D_m$,
$\Omega=\cup_{m=1}^M D_m$. We assume  that $D_m$
is a circle of radius $a$ centered at the point $\hat{x}_m$.

Let us
look for $u$ of the form \be\label{e29}
u(\hat{x})=u_0(\hat{x})+\sum_{m=1}^M\int_{S_m}g(\hat{x},t)\sigma_m(t)dt.
\ee We assume that the points $\hat{x}_m$ are distributed in a
bounded domain $D$ on the plane $P=xoy$ by formula \eqref{e2}. The
 field $u_0(\hat{x})$ is the same as in Section 2,
$u_0(\hat{x})=e^{i\kappa y}$,
and Green's function  $g$ is the same as in
formulas
\eqref{e16}-\eqref{e18}. It follows from \eqref{e2} that
$M=M(a)=O(\ln \frac{1}{a}).$ We define the effective field, acting
on the $D_j$ by the formula \be\label{e30}
u_e=u_e^{(j)}=u(\hat{x})-\int_{S_j}g(\hat{x},t)\sigma_j(t)dt,\quad
|\hat{x}-\hat{x}_j|>a,
\ee
which can also be written as
$$u_e(\hat{x})=u_0(\hat{x})+
\sum_{m=1, m\neq j}^M\int_{S_m}g(\hat{x},t)\sigma_m(t)dt.$$
We assume that the distance $d=d(a)$
between neighboring cylinders   is much greater
than $a$: \be\label{e31} d\gg a, \quad \lim_{a\to
0}\frac{a}{d(a)}=0. \ee

Let us rewrite \eqref{e29} as \be\label{e32} u=u_0+\sum_{m=1}^M
g(\hat{x},\hat{x}_m)Q_m+\sum_{m=1}^M\int_{S_m}[g(\hat{x},t)-g(\hat{x},\hat{x}_m)]\sigma_m(t)dt,
\ee where \be\label{e33} Q_m:=\int_{S_m}\sigma_m(t)dt. \ee As $a\to
0$, the second sum in \eqref{e32} (let us denote it $\Sigma_2$)
is negligible compared
with the first sum in \eqref{e32}, denoted $\Sigma_1$,
\be\label{e34}
|\Sigma_2|\ll
|\Sigma_1|,\quad a\to 0. \ee
The proof of this is similar to the one given in \cite{R509} for
a similar problem in $\R^3$.

Let us check that \be\label{e35}
|g(\hat{x},\hat{x}_m)Q_m|\gg
|\int_{S_m}[g(\hat{x},t)-g(\hat{x},x_m)]\sigma_m(t)dt|,\quad a\to 0.
\ee If $k|\hat{x}-\hat{x}_m|\gg 1$, and $k>0$ is fixed then
$$|g(\hat{x},\hat{x}_m)|=O(\frac{1}{|\hat{x}-\hat{x}_m|^{1/2}}), \qquad
|g(\hat{x},t)-g(\hat{x},x_m)|=O(\frac{a}{|\hat{x}-\hat{x}_m|^{1/2}}),$$
and $Q_m\neq 0$, so estimate \eqref{e35} holds.

If
$$|\hat{x}-\hat{x}_m|\sim
d\gg a,$$ then
$$|g(\hat{x},\hat{x}_m)|=O(\frac{1}{ln \frac{1}{a}}), \qquad
|g(\hat{x},t)-g(\hat{x},x_m)|=O(\frac{a}{d}),$$ as follows from the
asymptotics of $H_0^1(r)=O(\ln \frac{1}{r})$ as $r\to 0$, and
from the formulas $\frac{dH_0^1(r)}{dr}=-H_1^1(r)=O(\frac{1}{r})$ as $r\to
0$.
Thus,
\eqref{e35} holds for $|\hat{x}-\hat{x}_m|\gg d\gg a$.

Consequently, the
scattering problem is reduced to finding numbers $Q_m$, $1\leq m\leq
M$.

Let us estimate $Q_m$ asymptotically, as $a\to 0$. To do this,
we use the exact boundary condition on $S_m$, which yields
\be\label{e36} -u_e(s)=\int_{S_j}g(s,t)\sigma_j(t)dt,\quad s\in S_j.
\ee
The function  $u_e(s)$ is twice differentiable, so
$$u_e(s)=u_e(\hat{x}_j)(1+O(ka)).$$
Neglecting the term $O(ka)$ as $a\to
0$, rewrite equation \eqref{e36} as \be\label{e37}
-u_e(\hat{x}_j)=\int_{S_j}g(s,t)\sigma_j(t)dt. \ee This equation is
similar to \eqref{e19}: the role of $u_0(0)$ is played by
$u_e(x_j)$. Repeating the argument, given in Section 2, one obtains
a formula, similar to \eqref{e27}: \be\label{e38} Q_j=-\frac{2\pi
u_e(\hat{x}_j)}{\ln \frac{1}{a}}[1+o(1)],\quad a\to 0. \ee Formula,
similar to \eqref{e28}, is \be\label{e39}
u(\hat{x})\sim u_0(\hat{x})-\frac{2\pi}{\ln
\frac{1}{a}}\sum_{m=1}^Mg(\hat{x},\hat{x}_m)u_e(\hat{x}_m), \quad a\to 0.
\ee
The numbers $u_e(\hat{x}_m)$, $1\leq m\leq M$, in \eqref{e39} are not
known. Setting $\hat{x}=\hat{x}_j$ in \eqref{e39}, neglecting $o(1)$ term,
and using the
definition \eqref{e30} of the effective field, one gets a linear
algebraic system for finding numbers $u_e(\hat{x}_m)$: \be\label{e40}
u_e(\hat{x}_j)=u_0(\hat{x}_j)-\frac{2\pi}{\ln
\frac{1}{a}}\sum_{m\neq j}g(\hat{x}_j,\hat{x}_m)u_e(\hat{x}_m),\quad 1\le
j\le M. \ee
This system can be solved numerically if the number $M$ is not very
large, say $M\leq O(10^3)$.

If $M$ is very large, $M=M(a)\to \infty,\,\, a\to 0$,  then we derive
a linear
integral equation for the limiting effective field in the medium
obtained by embedding many cylinders.

Passing to the limit $a\to 0$ in system \eqref{e40} is done as
in \cite{R595}. Consider a partition of the domain $D$ into a union of
$\bf{P}$
small squares $\Delta_p$, of size $b=b(a)$, $b\gg d\gg a$. For
example, one may choose $b=O(a^{1/4})$, $d=O(a^{1/2})$, so that
there are many discs $D_m$ in the  square  $\Delta_p$. We assume that
 squares $\Delta_p$ and $\Delta_q$ do not have common interior points
if $p\neq q$. Let
$\hat{y}_p$ be
the center of $\Delta_p$. One can also choose as $\hat{y}_p$  any point
$\hat{x}_m$ in a domain
$D_m\subset \Delta_p$. Since $u_e$ is a continuous function, one may
approximate $u_e(\hat{x}_m)$ by $u_e(\hat{y}_p)$, provided that
$\hat{x}_m\subset \Delta_p$. The error of this approximation is
$o(1)$ as $a\to 0$. Let us rewrite the sum in \eqref{e40} as
follows:\be\label{e41} \frac{2\pi}{\ln\frac{1}{a}}\sum_{m\neq
j}g(\hat{x}_j,\hat{x}_m)u_e(\hat{x}_m)=2\pi
\sum^{\bf{P}}_{\stackrel{p=1}{x_j\notin
\Delta_p}}g(\hat{x}_j,\hat{y}_p)u_e(\hat{y}_p)\frac{1}{\ln
\frac{1}{a}}\sum_{x_m\in \Delta_p}1,\ee and use formula \eqref{e2}
in the form

\be\label{e42} \frac{1}{\ln \frac{1}{a}}\sum_{x_m\in
\Delta_p}1=N(\hat{y}_p)|\Delta_p|[1+o(1)]. \ee Here $|\Delta_p|$ is
the volume of the  square $\Delta_p$.

From \eqref{e41} and \eqref{e42}
one obtains: \be\label{e43} \frac{2\pi}{\ln \frac{1}{a}}\sum_{m\neq
j}g(\hat{x}_j,\hat{x}_m)u_e(\hat{x}_m)=
2\pi \sum^{\bf{P}}_{\stackrel{p=1}{\hat{x}_j\notin
\Delta_p}}
g(\hat{x}_j,\hat{y}_p)N(\hat{y}_p)u_e(\hat{y}_p)|\Delta_p|[1+o(1)].\ee
The sum in the right-hand side in \eqref{e43} is the Riemannian sum
for the integral \be\label{e44} \lim_{a\to
0}\sum_{p=1}^{\bf{P}}g(\hat{x}_j,\hat{y}_p)N(\hat{y}_p)u_e(\hat{y}_p)|\Delta_p|=\int_D
g(\hat{x},\hat{y})N(\hat{y})u(\hat{y})dy,\quad u(\hat{x})=\lim_{a\to
0}u_e(\hat{x}). \ee Therefore, system \eqref{e40} in the limit $a\to
0$ yields the integral equation for the limiting effective field
\be\label{e45}
u(\hat{x})=u_0(\hat{x})-2\pi\int_Dg(\hat{x},\hat{y})N(\hat{y})
u(\hat{y})d\hat{y}.
\ee
One obtains system \eqref{e40} if one solves equation
\eqref{e45} by a collocation method. Convergence of this method to
the unique solution of equation \eqref{e45} is proved in \cite{R563}.
Existence and uniqueness of the solution to equation \eqref{e45} are
proved as in \cite{R509}, where a three-dimensional analog of this
equation was studied.

Applying the operator $\Delta_2+\kappa^2$ to
equation \eqref{e45} yields the following differential equation for
$u(\hat{x})$: \be\label{e46}
\Delta_2u(\hat{x})+\kappa^2u(\hat{x})-2\pi N(\hat{x})u(\hat{x})=0.
\ee This is a Schr\"{o}dinger-type equation, and $u(\hat{x})$ is its
scattering solution corresponding to the incident wave
$u_0=e^{i\kappa y}$.

Let us assume that $N(x)=N$ is a constant. One concludes from \eqref{e46}
that the limiting medium, obtained by
embedding many perfectly conducting circular cylinders, has
new parameter $\kappa_N^2:=\kappa^2-2\pi N$. This means
that $k^2=\kappa^2 +k_3^2$ is replaced by $\tilde{k}^2:=k^2-2\pi N$.
The quantity $k_3^2$ is not changed. One has $\tilde{k}^2=\omega^2n^2$,
$k^2=\omega^2 n_0^2$. Consequently, $n^2/n_0^2= (k^2-2\pi N)/k^2$.
Therefore, the  new
refraction coefficient $n^2$ is
\be\label{e47} n^2=n_0^2(1-2\pi N k^{-2}),
\ee
Since the number $N>0$ is at our
disposal, equation \eqref{e47} shows that choosing suitable
$N$ one can create a medium with a smaller, than $n_0^2$, refraction
coefficient.

In practice  one does not go to the limit $a\to 0$, but
chooses a sufficiently small $a$. As a result, one obtains a medium
with a refraction coefficient $n^2_a$, which differs from
\eqref{e47} a little, $\lim_{a\to 0} n_a^2=n^2.$

\section{Conclusions}
Asymptotic, as $a\to 0$, solution is given of the EM wave scattering
problem by many perfectly conducting parallel cylinders of radius
$a$. The equation for the effective field in the limiting medium
obtained when  $a\to 0$ and the distribution of the embedded cylinders
is given by formula \eqref{e2}. The presented theory gives formula
\eqref{e47}
for the refraction coefficient in the limiting  medium. This formula shows
how the distribution of the cylinders
influences the refraction coefficient.

\section{Appendix}
Let us derive formulas \eqref{e14}-\eqref{e15}. Look for the
solution to \eqref{e3}-\eqref{e4} of the form: \be\label{eA1}
E_1=e^{ik_3z}\tilde{E}_1(x,y),\,\,\,
E_2=e^{ik_3z}\tilde{E}_2(x,y),\,\,\, E_3=e^{ik_3z}u(x,y), \ee
\be\label{eA2} H_1=e^{ik_3z}\tilde{H}_1(x,y),\,\,\,
H_2=e^{ik_3z}\tilde{H}_2(x,y),\,\,\, H_3=0, \ee where $k_3=const$.
Equation \eqref{e3} yields \be\label{eA3}
u_y-ik_3\tilde{E}_2=i\omega \mu \tilde{H}_1,\,\,\,
-u_x+ik_3\tilde{E}_1=i\omega \mu \tilde{H}_2,\,\,\,
\tilde{E}_{2,x}=\tilde{E}_{1,y}, \ee where, e.g.,
$\tilde{E}_{j,x}:=\frac{\partial \tilde{E}_{j}}{\partial x}$.
Equation \eqref{e4} yields \be\label{eA4} ik_3\tilde{H}_2=i\omega
\epsilon \tilde{E}_1,\,\,\, ik_3\tilde{H}_1=-i\omega \epsilon
\tilde{E}_2,\,\,\, \tilde{H}_{2,x}-\tilde{H}_{1,y}=-i\omega \epsilon
u. \ee Excluding $\tilde{H}_j$, $j=1,2$, from \eqref{eA3} and using
\eqref{eA4}, one gets \be\label{eA5}
\tilde{E}_1=\frac{ik_3}{\kappa^2}u_x,\,\,\,
\tilde{E}_2=\frac{ik_3}{\kappa^2}u_y,\,\,\, \tilde{E}_3=u, \ee
\be\label{eA6} \tilde{H}_1=\frac{k^2u_y}{i\omega
\mu \kappa^2},\,\,\, \tilde{H}_2=-\frac{k^2u_x}{i\omega
\mu \kappa^2}u_x,\,\,\, \tilde{H}_3=0. \ee Since
$E_j=\tilde{E}_j e^{ik_3z}$ and $H_j=\tilde{H}_j e^{ik_3z}$,
formulas \eqref{e14}-\eqref{e15} follow immediately from
\eqref{eA5}-\eqref{eA6}.

\end{document}